\documentclass[twocolumn,showpacs,final,prd]{revtex4}
\usepackage{graphicx}
\newcommand{\be}{\begin{equation}}
\newcommand{\ee}{\end{equation}}
\newcommand{\bea}{\begin{eqnarray}}
\newcommand{\eea}{\end{eqnarray}}

\newcommand{\stgg}{$\Gamma_{\sigma\to\gamma\gamma}$}
\newcommand{\stggqq}{$\Gamma^{q\bar{q}}_{\sigma\to\gamma\gamma}$}
\newcommand{\qllsm}{QLL$\sigma$M}
\begin{document}
\title{Comment on ``Two-photon decay of the sigma meson''}
\author{Eef van Beveren}
\email{eef@teor.fis.uc.pt}
\affiliation{Centro de F\'{\i}sica Te\'{o}rica,
Departamento de F\'{\i}sica, Universidade de Coimbra,
P-3004-516 Coimbra, Portugal}
\author{Frieder Kleefeld}
\email{kleefeld@cfif.ist.utl.pt}
\author{George Rupp}
\email{george@ist.utl.pt}
\affiliation{Centro de F\'{\i}sica das Interac\c{c}\~{o}es Fundamentais,
Instituto Superior T\'{e}cnico, Technical University of Lisbon,
P-1049-001 Lisboa Codex, Portugal}
\author{Michael D.\ Scadron}
\affiliation{Department of Physics, University of Arizona, Tucson,
AZ 85721, USA}
\email{scadron@physics.arizona.edu}
\date{\today}

\begin{abstract}
We comment on a recent paper by Giacosa, Gutsche, and Lyobovitskij, in
which it is argued that a quarkonium interpretation of the $\sigma$ meson
should give rise to a much smaller two-photon decay width than commonly
assumed. The reason for this claimed discrepancy is a term in the
transition amplitude, necessary for gauge invariance, which allegedly is
often omitted in the literature, including the work of the present authors.
Here we show their claims to be incorrect by demonstrating, in the context of
the Quark-Level Linear $\sigma$ Model, that the recently extracted
experimental value $\Gamma_{\sigma\to2\gamma}=(4.1\pm0.3)$ keV is 
compatible with a $q\bar{q}$ assignment for the $\sigma$, provided that meson
loops are taken into account as well.
\end{abstract}


\maketitle

\section{Introduction}
In a recent paper \cite{GGL08}, Giacosa, Gutsche, and Lyubovitskij (GGL)
studied the two-photon decay width of the $\sigma$ meson, alias $f_0(600)$
\cite{PDG06}, based on the presupposition that it is a $q\bar{q}$ state. They
employed two simple perturbative sigma models, one purely local, comprising
$\sigma$, $\pi$, quark and antiquark fields, and the other nonlocal, with only 
$\sigma$, $q$, and $\bar{q}$, besides an extended covariant vertex function.
The principal result of their work was that, in contrast with what is generally
assumed, a $q\bar{q}$ assignment for the $\sigma$ should lead to a width
$\Gamma_{\sigma\to\gamma\gamma}$ much smaller than the recently reported value
of $(4.1\pm0.3)$ keV resulting from an analysis by Pennington \cite{P06}, as
well as the 3 values given in the 2006 PDG tables \cite{PDG06}, and probably
even less than 1 keV. Therefore, GGL concluded that, if the large experimental
$\gamma\gamma$ width is confirmed, a quarkonium interpretation of the $\sigma$
is not favored, \em ``contrary to usual belief.'' \em As an explanation for
their very small \stgg\ prediction, GGL argued that a term in the
quark-triangle loop diagram, necessary for gauge invariance, largely cancels
the lead term, thus resulting in a small total amplitude. Moreover, GGL claimed
that the former term is \em ''often neglected'', \em including in previous work
of ours and our co-authors \cite{BKRS02,KBRS02,DLS99,SKR06}.

In this Comment, we shall show that GGL are mistaken on several points. First
of all, we have \em not \em \/unduly neglected any term in the evaluation of
the quark triangle diagram in Refs.~\cite{BKRS02,KBRS02,DLS99,SKR06}. When we
disregarded the term in question, this was fully justified, since the term was
zero or negligible. Secondly, the small \stgg\ value obtained by GGL is a
consequence of a very low $\sigma$ mass, in combination with a relatively large
constituent
quark mass, at least in the local case. For the nonlocal Lagrangian, their tiny
\stgg\ value is rather an indication for the inadequacy of the Lagrangian
itself. Thirdly, we demonstrate, by explicit calculation, how important
meson-loop contributions are, which is in principle admitted by GGL, but not
concretized.

In Sec.~\ref{sigma} of this Comment, we study in detail the
two-photon width of the $\sigma$ meson, in the context of the quark-level
linear $\sigma$ model (QLL$\sigma$M) \cite{DS95}, showing that a good agreement
with data is achieved. In Sec.~\ref{cncls} we present our conclusions.

\section{Two-photon width of the $\sigma$ in the QLL$\sigma$M}
\label{sigma}
Given the scalar amplitude structure \cite{DLS99,KBRS02,DEMSB94}
${\cal M}\epsilon_\nu(k')\epsilon_\mu(k)
\left(g^{\mu\nu}k'\cdot k-k'^\mu k^\nu\right)$,
the rate for the decay of a scalar meson $S$ into two photons reads
\be
\Gamma(S\to\gamma\gamma)=\frac{m^3_S|{\cal M}_{S\to\gamma\gamma}|^2}{64\pi}\;.
\label{stogg}
\ee
If one assumes, as GGL do, that the $\sigma$ is a scalar $q\bar{q}$ state, then
the principal contribution to the amplitude ${\cal M}_{\sigma\to\gamma\gamma}$
comes from the up and down quark triangle diagrams (see e.g.\ FIG.~1 in
Ref.~\cite{GGL08}), yielding (with $N_c=3$) \\[-4mm]
\be
{\cal M}^{n\bar{n}}_{\sigma\to\gamma\gamma} \; = \;
\frac{5\alpha}{3\pi f_\pi} 2\xi_n[2+(1-4\xi_n)I(\xi_n)] \; ,
\label{mnsig}
\ee
where $\alpha=e^2/4\pi$, $\xi_n=m_n^2/m_\sigma^2$ ($n$ stands for $u$ or $d$),
and $I(\xi)$ is the triangle loop integral given by
\be
I(\xi) \left\{
\begin{array}{lll}
=\;\displaystyle\frac{\pi^2}{2}-2\log^2\left[\sqrt{\frac{1}{4\xi}}+
\sqrt{\frac{1}{4\xi}-1}\:\right] + \\[4mm] \displaystyle\;\;\;\;\;\;
2\pi i\log\left[\sqrt{\frac{1}{4\xi}}+\sqrt{\frac{1}{4\xi}-1}\:\right]
\;(\xi\leq0.25)\:,\\[5mm]
=\;\displaystyle2\arcsin^2\left[\sqrt{\frac{1}{4\xi}}\:\right]
\;\;(\xi\geq0.25)\;.
\end{array}
\right.
\label{ixi}
\ee
These Eqs.~(\ref{mnsig}) and (\ref{ixi}) exactly correspond to Eqs.~(2)
and (4) in Ref.~\cite{GGL08}, with the proviso that GGL
defined the $\sigma$-$\bar{q}$-$q$ coupling in their Lagrangian as
$g_\sigma/\sqrt{2}$ instead of our \qllsm\ coupling $g$, the latter being
related to $f_\pi$ above via the Goldberger-Treiman relation $m_q=f_\pi g$
\cite{BKRS02,KBRS02,DLS99,SKR06}.
Ignoring for the moment possible meson-loop contributions as well as an
$s\bar{s}$ component in the $\sigma$, we can use Eq.~(\ref{mnsig}) to calculate
\stgg, for different $\sigma$ and quark masses. Also, we can check what the
importance is of the term involving $I(\xi)$.

However, let us first deal with the allegation by GGL that we had erroneously
neglected the $I(\xi)$ term in previous work. Well, in Ref.~\cite{BKRS02}
we simply worked in the, perfectly well-defined, Nambu--Jona-Lasinio (NJL)
\cite{NJL61} limit ($m_\sigma=2m_q$) of the \qllsm, in which the term in
question vanishes
identically, using quite reasonable $\sigma$ and quark masses of 630~MeV
and 315~MeV, respectively. The resulting \stgg, ignoring meson-loops,
would then be 2.18~keV. But accounting for an estimate of the pion-loop
contribution as well yielded the prediction of 3.76 keV \cite{BKRS02}, in good
agreement with experiment, then and now. In Ref.~\cite{SKR06}, Eq.~(101),
again the NJL limit of the \qllsm\ was used, but now also including an estimate
for the kaon loop, besides the pion loop, leading to a slightly smaller 
result, but still very much larger than any of GGL's predictions
(also see Ref.~\cite{SRKB04}). Finally,
in Refs.~\cite{KBRS02,DLS99} \stgg\ was not even considered, thus making the
critique by GGL completely void. Moreover, note that in Ref.~\cite{KBRS02}
we did use the full expressions of Eqs.~(\ref{mnsig}) and (\ref{ixi}) above
when necessary, namely in the case of the $f_0$(1370) meson.

Let us now carry out a more detailed analysis of \stgg\ in a \qllsm\ setting,
employing Eqs.~(\ref{mnsig}) and (\ref{ixi}).
Working beyond the chiral limit (CL), we may take the NJL value
$m_\sigma=675$~MeV for $m_n=337.5$ MeV \cite{S08}, where $m_n$ stands for the
nonstrange (up or down) quark mass. Still neglecting $n\bar{n}$-$s\bar{s}$
mixing and meson loops, this gives \stggqq$=2.68$~keV. Taking a somewhat more
realistic value of $m_\sigma=666$~MeV \cite{S08}, away from the CL, the latter
width gets reduced to 2.44 keV. If we now also allow for the admixture of a
small $s\bar{s}$ component in the $\sigma$, with a nonstrange-strange mixing
angle of, say, $-10.1^\circ$ \cite{S08}, then we get \stggqq$=2.49$~keV, for
the often used \cite{SKR06} \qllsm\ quark masses $m_n=337.5$~MeV and
$m_s=486$~MeV. Note that this $s\bar{s}$ component, with amplitude
\be
{\cal M}^{s\bar{s}}_{\sigma\to\gamma\gamma} \; = \; \frac{\sqrt{2}\alpha g}
{3\pi m_s}2\xi_s[2+(1-4\xi_s)I(\xi_s)]\;,
\label{mssig}
\ee
contributes with a weight factor of only $\sqrt{2}\alpha m_n/3\pi f_\pi m_s$
(using the GT relation $m_n=f_\pi g$), as compared to
$5\alpha /3\pi f_\pi$ from Eq.~(\ref{mnsig}) in the $n\bar{n}$ case,
since the charge of a strange quark is $-1/3$ \cite{KBRS02}.

Next we are going to add meson-loop contributions as well. Now, in the
framework of the \qllsm, loops with charged mesons that couple to the $\sigma$
include those with pions and kaons, as well as those with the scalar mesons
$\kappa$(800) and $a_0$(980). 
The expression for a gauge-invariant meson-loop contribution to the
two-photon amplitude mainly differs from the quark triangle in
Eq.~(\ref{mnsig}) because of the presence of a seagull graph (see e.g.\
Ref.~\cite{DEMSB94}, first paper), yielding a total amplitude
\be
{\cal M}^{MM}_{\sigma\to\gamma\gamma}\;=\;-\frac{2g'\alpha}{\pi m^2_M}
\left[-\frac{1}{2}+\xi I(\xi)\right] \; , \;\;\;
\xi=\frac{m^2_M}{m^2_\sigma} \; ,
\label{mloop}
\ee
where the minus sign stems from the opposite statistics with respect to the
quark-loop case, and $g'$ is the cubic \qllsm\ meson coupling. For the meson
loops pertinent to the $\sigma$, we shall need the 3-meson couplings
\cite{DS95,KBRS02,SKR06}
\be
\begin{array}{lcl}
g_{\sigma_{n\bar{n}},\pi\pi} & = & \displaystyle\frac{\cos^2(\phi_S)m^2_\sigma+
              \sin^2(\phi_S)m^2_{f_0(980)}-m^2_{\pi^\pm}}{2f_\pi}\;, \\[3mm]
g_{\sigma_{s\bar{s}},\pi\pi} & = & 0 \;, \\[3mm]
g_{\sigma_{n\bar{n}},KK}     & = & \displaystyle\frac{\cos^2(\phi_S)m^2_\sigma+
              \sin^2(\phi_S)m^2_{f_0(980)}-m^2_{K^\pm}}{2f_K}\; , \\[4mm]
g_{\sigma_{s\bar{s}},KK}     & = & \displaystyle\frac{\sin^2(\phi_S)m^2_\sigma+
             \cos^2(\phi_S)m^2_{f_0(980)}-m^2_{K^\pm}}{\sqrt{2}\,f_K}\;,\\[4mm]
g_{\sigma_{n\bar{n}},\kappa\kappa} & = & \displaystyle\frac{\cos^2(\phi_S)
m^2_\sigma+\sin^2(\phi_S)m^2_{f_0(980)}-m^2_{\kappa}}{2\,(f_\pi-f_K)}\;,\\[4mm]
g_{\sigma_{s\bar{s}},\kappa\kappa} & = & \displaystyle\frac{\sin^2(\phi_S)
m^2_\sigma+\cos^2(\phi_S)m^2_{f_0(980)}-m^2_{\kappa}}{\sqrt{2}\,(f_K-f_\pi)}\;,
\\[4mm]
g_{\sigma_{n\bar{n}},a_0a_0} & = & 3g_{\sigma_{n\bar{n}},\pi\pi} \;, \\[3mm]
g_{\sigma_{s\bar{s}},a_0a_0} & = & 0 \;, 
\end{array}
\label{cubic}
\ee
where $\phi_S$ is the scalar mixing angle, and
$f_K=f_\pi\,(m_s/m_n+1)/2\approx1.22\,f_\pi$. The cubic coupling of the
physical $\sigma$ meson to the three channels is then given by
\be
g'_{\sigma,MM} \; = \; \cos(\phi_S)g_{\sigma_{n\bar{n}},MM} -
                       \sin(\phi_S)g_{\sigma_{s\bar{s}},MM} \; .
\label{sigmamm}
\ee
Note that we neglect here small OZI-violating corrections to the \qllsm\
three-meson couplings, just as in previous work of ours \cite{KBRS02}.
Such contributions will be included in a forthcoming study.

Now we are in a position to do a complete calculation of \stgg, with both quark
and meson loops accounted for. Note that the imaginary part of $I(\xi$), as
given by the $\xi<0.25$ case in Eq.~(\ref{ixi}), will be included for the
pion-loop amplitude. If we choose again a scalar mixing angle of $-10.1^\circ$
and take $m_{\kappa}=800$~MeV, we obtain a total two-gamma width
\be
\Gamma_{\sigma\to\gamma\gamma}^{q\bar{q}+MM} \; = \; 3.50 \; \mbox{keV} \; .
\label{gstgg}
\ee
This rate corresponds to a total amplitude modulus
$|{\cal M}|=4.88\times10^{-2}$ GeV$^{-1}$, which can be decomposed in terms of
the partial quark- and meson-loop amplitudes
\be
\begin{array}{rcl}
\Re\mbox{e}\,{\cal M}_{n\bar{n}}&=&4.01\times10^{-2}\;\mbox{GeV}^{-1}\;,\\[1mm]
\Re\mbox{e}\,{\cal M}_{s\bar{s}}&=&1.09\times10^{-2}\;\mbox{GeV}^{-1}\;,\\[1mm]
{\cal M}_{\pi\pi}&=&(1.19 - i 1.03) \times 10^{-2}\;\mbox{GeV}^{-1}\;,\\[1mm]
{\cal M}_{KK}      & = & -1.83 \times 10^{-3} \; \mbox{GeV}^{-1} \; , \\[1mm]
{\cal M}_{\kappa\kappa} &=&-2.06 \times 10^{-3} \; \mbox{GeV}^{-1}\;,\\[1mm]
{\cal M}_{a_0a_0}  & = & -1.50 \times 10^{-3} \; \mbox{GeV}^{-1} \; .
\end{array}
\label{amplitudes}
\ee
Note that here the relative sign between quark and meson loops has already
been included. Also observe that the kaon, $\kappa$, and $a_0(980)$ loops
reduce the contribution of the pion loop, so that the net effect of the
meson loops on the two-photon width is about $+40\%$.

Taking a somewhat more negative value for the scalar mixing angle, e.g.\
$\phi_S=-18^\circ$ \cite{SKR06}, only reduces the total two-phton width to
3.39~keV. This prediction as well as the former one are fully compatible with
the corresponding PDG \cite{PDG06} data, and also not at odds with Pennington's
recent result \cite{P06}.

In contrast, the sensitivity of \stgg\ to the $\sigma$ mass is much stronger,
which is obvious from Eq.~(\ref{stogg}), relating width and amplitude via
$m_\sigma$ cubed. This can also by seen in FIG.~2 of the paper \cite{GGL08}
by GGL themselves, where e.g.\ an $m_\sigma$ of 650 MeV, with $m_q=350$ MeV,
would yield a \stggqq\ of roughly 2.5~keV, in good agreement with our value
of 2.44~keV above. However, by taking a very small $m_\sigma$ of 440~MeV, as
GGL choose to do, one obtains a much smaller \stgg, even when meson loops
are included. For instance, if we assume the $\sigma$ to be purely $n\bar{n}$
and take $m_q=250$~MeV, \stgg\ becomes 0.67~keV, even with the 3 meson-loop
contributions included, which should be compared to GGL's value of 0.54~keV
(see TABLE I of Ref.~\cite{GGL08}) for the pure $q\bar{q}$ case. Neglecting
in this scenario the term proportional to $I(\xi)$ would indeed increase our
result of 0.67~keV to 1.38~keV, but this is of course an error we have not
and will not make.

At this point, we also take exception at GGL's claim \em
``\ldots the results for \stgg\ at a fixed pole mass of $M_\sigma=440$~MeV as
favored by recent theoretical and experimental works [16,20]'', \em where
their reference no.~20 is our Ref.~\cite{PDG06}, i.e., the 2006 PDG Review of
Particle Physics. It is simply false to state that the PDG favors a $\sigma$
pole mass of 440~MeV. The truth is that the PDG listings mention ``{\bf
(400--1200)\boldmath{$-i$}(250--500) OUR ESTIMATE}'', for the $f_0(600)$
$T$-matrix pole (i.e., $S$-matrix pole) as a function of $\sqrt{s}$.
On the other hand, the theoretical papers referred
to by GGL include the Roy-equation analysis by Caprini, Colangelo, and
Leutwyler \cite{CCL06}, which indeed found 441~MeV for the real part of the
$\sigma$ $S$-matrix pole, besides an imaginary part of 272~MeV. However, it is
a common mistake to confuse the real part of the pole with the `mass' of a
broad resonance, especially when the resonance is certainly not of a pure
BW type, like e.g.\ the $\sigma$, which is strongly distorted due to the
$\pi\pi$ threshold and the Adler zero not far below \cite{B03}. Notice
that, in the latter analysis, the `mass' of the $\sigma$ at which the $\pi\pi$
phase shift passes through $90^\circ$ --- by definition the $K$-matrix pole ---
lies at 926~MeV. This does not mean that this \em is \em \/the $\sigma$ mass,
but just demonstrates the difficulty of assigning \em any \em \/specific mass
to a broad non-BW resonance. Anyhow, our above choice of 666~MeV, in the
context of the \qllsm, is surely more reasonable than naively taking the
real part of a pole that is already significantly lower than the `world
average' \cite{PDG06,eef} of $\sigma$ poles. 

To conclude this section, we note that the $Z=0$ compositeness condition,
discussed by GGL in the context of their nonlocal Lagrangian,
is manifestly satisfied in the --- nonperturbative and selfconsistent ---
\qllsm, provided $\xi=m^2_q/m^2_\sigma\leq0.25$, with $g_\sigma$ \em  not
\em \/depending on $m_\sigma$. \\

\section{Conclusions}
\label{cncls}
In the present Comment we have shown that GGL incorrectly referred to and
criticized our previous papers on the subject. Moreover, we have demonstrated,
via an explicit and detailed calculation in the context of the \qllsm, that
the reported experimental values of \stgg\ give quantitative support to a
$q\bar{q}$ interpretation of the $\sigma$ meson, provided that one uses a
reasonable $\sigma$ mass and also includes meson-loop contributions, besides
the quark loop considered by GGL.

Finally, let us comment on the nonlocal Lagrangian employed by GGL besides the
local one. Their justification was: \em ``However, the local approach is no
longer applicable for values of $M_\sigma$ close to threshold, as will be 
evident from the discussion of the next section.'' \em Well, as already
mentioned above, the \qllsm\ is a \em local \em \/renormalizable field theory,
exactly satisfying the $Z=0$ compositeness condition close to --- but below ---
threshold, due to its nonperturbative and selfconsistent formulation
\cite{DS95}. This condition can be rigorously described in both the \qllsm\ 
and the NJL model, in terms of a log-divergent gap equation \cite{S98}. The
latter can also be expressed via a four-dimensional ultraviolet cutoff
$\Lambda$, resulting in a value $\Lambda\approx2.3m_q$. For a nonstrange quark
mass of 337.5 MeV, this gives $\Lambda\approx750$ MeV, which is an energy scale
that clearly separates the `elementary' $\sigma$ from e.g.\ the
'composite' $\rho$ meson. For further details, we refer to Ref.~\cite{S98}.

In contrast, GGL were probably thinking in perturbative terms when
going from their local $\sigma$-model Lagrangian to the nonlocal case. In view
of the numerical results of the latter model, which produces even tinier values
for \stgg\ than their local approach, we are led to conclude that Nature
rather disfavors a nonlocal realization of chiral symmetry than a $q\bar{q}$
interpretation of the $\sigma$ meson.

This work was supported by the {\it Funda\c{c}\~{a}o para a
Ci\^{e}ncia e a Tecnologia} \/of the {\it Minist\'{e}rio da Ci\^{e}ncia,
Tecnologia e Ensino Superior} \/of Portugal, under contract
POCI/FP/81913/2007.

\end{document}